\documentstyle[twoside,psfig,fleqn,espcrc2]{article}

\newcommand{\AmS}{{\protect\the\textfont2
  A\kern-.1667em\lower.5ex\hbox{M}\kern-.125emS}}
\def\gam{$\gamma$}
 
\title{ Intensive monitoring of the strongly variable BL~Lac S5 0716+714}

\author{K.\ Otterbein\address{Landessternwarte Heidelberg-K\"onigstuhl, 
			       D-69117 Heidelberg, Germany},
        M.\ J.\ Hardcastle\address{Department of Physics, University of 
				   Bristol, Tyndall Avenue, Bristol BS8 1TL, 
				   UK},
        S.\ J.\ Wagner$\rm ^a$
        and 
        D.\ M.\ Worrall$\rm ^{b, }$\address{Harvard-Smithsonian Center for 
					Astrophysics, Cambridge, MA 02138}}
       
\begin{document}

\begin{abstract}
The BL~Lac object S5 0716+714 was monitored during a multifrequency campaign in
1996. Preliminary analysis of
the optical, ROSAT and RXTE data are presented. Strong variability on short
time scales was observed. The data suggest an interpretation within a
multi-component model.
\end{abstract}

\maketitle

  \section{Introduction}
Variability studies of AGN, in particular of the extreme blazar class, have 
proven to be powerful tools for the investigation of the nature of these 
objects. Shortly after the discovery of quasars the observation of variability
of their radio fluxes on time scales of a few months \cite{sho65,den65} led to 
the inference of relativistic bulk motion \cite{ree66,wol66} and consequently 
fuelled the development of the relativistic jet model \cite{bla78}. 
Short term variability on time scales of a day (intraday variability, IDV see
review \cite{wag95}) provides even stronger constraints on the source size and
the physical conditions.
Correlations between IDV flares seen in different bands, particularly 
in the optical, UV, X-rays \cite{ede95,wag96}, provide the most conclusive
implications for physical models.

Despite the success of the relativistic jet model,
the emission processes which produce the overall spectral energy 
distribution in blazars are still unknown. The smooth radio to UV continuum is 
generally agreed to be synchrotron emission from the relativistic jet. However
the origin of the X-ray and \gam -ray emission is far from clear. The majority
of present-day models explain the high-energy emission by soft seed 
photons which get up-scattered by the relativistic electrons of the jet plasma. 
Thus the main question is where do the soft seed photons originate? Since most 
of the total power emitted in blazars is in X-rays and \gam -rays, solving this 
problem is crucial for our understanding of blazars.

In order to understand the emission processes in blazars, multifrequency 
monitoring of the fluxes is necessary to trace the evolution of
variability patterns throughout the spectrum and to determine reliable time 
lags. The lags serve as diagnostics of the emission processes and allow one 
to infer the structure of the relativistic jet.

  \section{The case of S5\,0716+714}
The source S5 0716+714\footnote{We will drop the designation S5 for the rest of 
the paper} shows a featureless optical spectrum and its optical 
continuum flux 
is significantly polarized \cite{bie81}. Due to the featureless
spectrum the redshift of the object is unknown. A lower limit of $\rm z \ge
0.25$ is inferred from the absence of a host galaxy brighter than
$\rm M_{R} = -21.0$ in deep optical images \cite{wag96}. 

The source belongs to a well studied subsample 
of the S5 catalogue \cite{kue81} which is selected by its high radio 
flux ($\rm S_{5\,GHz} \ge 1\,Jy$), by its flat radio spectrum ($\alpha \ge 
-0.5, S_{\nu} \sim \nu^{\alpha}$) and by its high declination 
($\delta \ge 70^{\circ}$) \cite{eck86}. 0716+714 displays a one sided VLBI 
jet which strongly bends towards the position angle of the arcsecond-scale 
structure \cite{eck87}. It is one of the first sources with reported intra-day 
variability (IDV). 0716+714 is also detected by ROSAT, SAX, and EGRET 
and has shown X-ray and \gam -ray variability \cite{wag96,chi97}. 

  \begin{figure}[t]
  \mbox{}\\[3mm]
  \psfig{figure=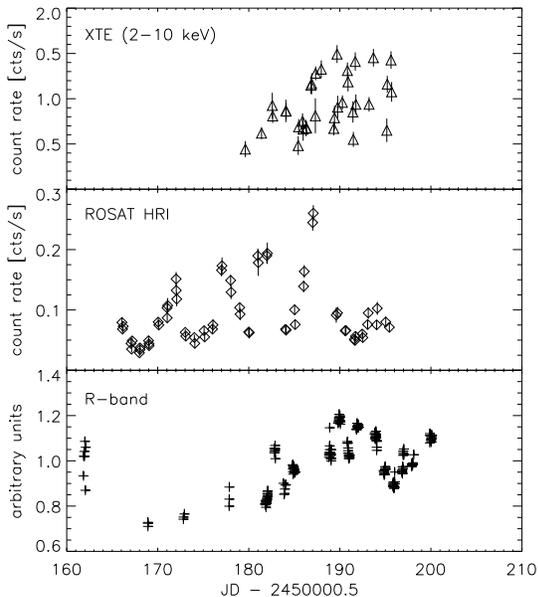,width=7.1cm,angle=90}
  \caption[]{\label{fg:lcrv_all} Light curves of the BL~Lac 
   0716+714 in different energy bands during the March/April multifrequency 
   campaign in 1996. From top to bottom: RXTE band ($2-10$\,keV),
   ROSAT HRI ($0.1-2.4$\,keV) and optical R-band.}
  \end{figure}
Flux variability of 0716+714 has been seen in almost all observations carried 
out to date regardless of wavelength. The variability time scales observed 
range from several hours to days. Correlated variability between the radio and 
optical regimes was found in February 1990 (\cite{wag96} and references 
therein). Despite a sparse sampling of the data, simultaneous optical and X-ray 
measurements during a 21\,ksec ROSAT PSPC observation in March 1991 suggest a 
close correspondence between these spectral bands. A ROSAT X-ray spectral index 
of $\rm \alpha_x = - 1.9\pm 0.2$ and a two-point optical to X-ray spectral index
of $\rm \alpha_{ox} = -1.7$ imply that the X-ray emission at 1\,keV 
may still be synchrotron emission. A more detailed review of the results of 
previous multifrequency campaigns is given in \cite{wag96}. 

\vspace{-1mm}
  \section{Multifrequency campaign in 1996}
In spring 1996 a multifrequency campaign ranging from the mm-regime to the
EGRET \gam -ray regime was organized. In this contribution we present 
preliminary results of the optical and X-ray data analysis. We focus on the
results obtained with the ROSAT and RXTE satellites, noting that because of the 
low count rate measured with the RXTE, these results are particularly sensitive 
to the RXTE background modelling which continues to be refined.

  \subsection{Data}
Figure \ref{fg:lcrv_all} shows the light curves of 0716+714 in three different 
energy bands. The top panel shows the data obtained with the RXTE-satellite, 
which pointed at 0716+714 for 34 individual observations during April 6$^{\rm 
th}$ until April 22$^{\rm nd}$. An average count rate over the $2-10$\,keV band 
of 0.95\,cts/sec was measured. The pointings lasted 2.8\,ksec on average.

ROSAT pointed at the source 28 times during the period from March 
24$^{\rm th}$ until April 22$^{\rm nd}$. A nearly daily sampling was achieved. 
The average pointing lasted 2.8\,ksec and the average count rate was 
0.1\,cts/sec. Several of these daily pointings had large gaps and stretched 
over as long as half a day.

The optical light curve is derived from measurements made with the
70\,cm telescope at the Heidelberg site and kindly provided by H.\ Bock. 
The complete optical data set including observations performed at many other 
sites will be discussed elsewhere.

  \section{Results}
  \begin{figure}[h]
  \psfig{figure=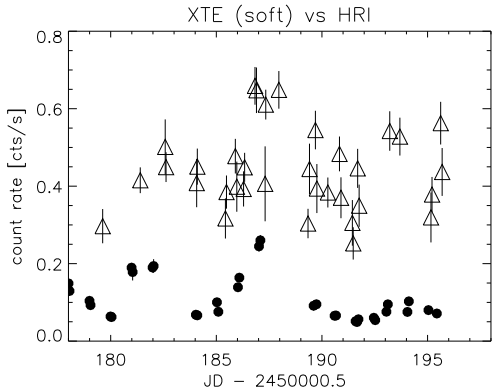,width=6.8cm,angle=90}
  \caption{\label{fg:comp-soft} Light curve of 0716+714 in
      the soft RXTE band ($2-4$\,keV, open triangles) compared to the ROSAT HRI
      ($0.1-2.4$\,keV, filled circles) count rate variations. The RXTE data is 
      shifted by 0.1\,cts/s to the top.}
  \end{figure}
  \begin{figure}
  \psfig{figure=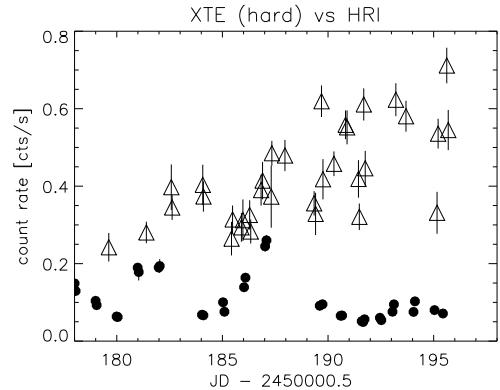,width=6.8cm,angle=90}
  \caption{\label{fg:comp-hard} Light curve of 0716+714 in 
   the hard RXTE band ($4-10$\,keV, open triangles) compared to the ROSAT HRI 
   ($0.1-2.4$\,keV, filled circles) count rate variations. The RXTE count rate 
   is divided by 2 and shifted by 0.15\,cts/s to the top.}
  \end{figure}
0716+714 showed strong and rapid X-ray and optical variations during the 
campaign (Figure \ref{fg:lcrv_all}). Indications of variations on time 
scales of several hours were found in all bands. Due to under-sampling in 
the X-ray bands, especially in the ROSAT band, this kind of variability is most 
obvious in the optical. Surprisingly, the variations seen in the three bands 
are not correlated. The apparent discrepancy between the behaviour in adjacent 
RXTE and ROSAT bands is unexpected.

In order to investigate the discrepancy between RXTE and ROSAT, the
total RXTE band was split into two distinct bands: a soft band ($2-4$\,keV) and
a hard band ($4-10$\,keV). Figure \ref{fg:comp-soft} shows the soft RXTE data 
(open triangles) compared to the ROSAT 
HRI data (filled circles). The figure suggests a better overall agreement with 
the ROSAT data than for the broad-band RXTE data. Figure \ref{fg:comp-hard} 
shows that the hard RXTE data (open triangles) do not
follow the variability characteristics of the ROSAT data, and here the
gradually rising baseline is particularly suggestive of the trend seen
in the optical data.  Thus we conclude that the X-ray data suggest
two components with different variability characteristics; one
component is dominant in soft X-rays and the other at energies
greater than about 4\,keV.

  \begin{figure}[t]
  \mbox{}\\[3mm]
  \psfig{figure=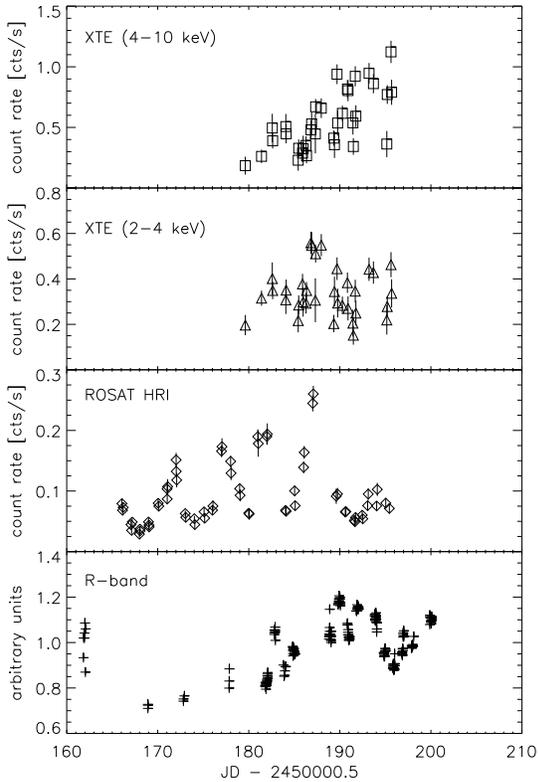,width=7.1cm,angle=90}
  \caption{\label{fg:lcrv_split} Light curves of 0716+714 in 1996.
   for various spectral bands: hard RXTE ($4-10$\,keV), soft XTE ($2-4$\,keV), 
   ROSAT HRI ($0.1-2.4$\,keV) and optical R-band.}
  \end{figure}
  \section{Discussion}
It is evident from Figure \ref{fg:lcrv_all} that the trends in the ROSAT band 
and in the optical band are not correlated.
Regarding the correspondence of the soft X-ray band and the optical band 
suggested by the previous ROSAT observation in 1991 the source seems to have 
changed its variability behaviour. Despite an apparent mismatch of the optical 
and the soft X-ray light curves the sampling of the optical data is at present 
insufficient for a more detailed investigation. Especially the most prominent 
features seen in the ROSAT band are not sampled in the optical data presented. 

The distinct variability properties of the soft and the hard RXTE band might 
indicate two source regions. The spectral characteristics of the source as 
observed with RXTE also suggest the presence of more than one spectral 
component. The slope of a power law model fitted to the total RXTE band is 
$\rm \alpha_{tot} = -2.1 \pm 0.1$ whilst the spectral index of the hard band for
a simple power law model is $\rm \alpha_{4-10\,keV} = -1.5 \pm 0.15$. The 
flattening of the spectrum in the hard RXTE band suggests the presence of at 
least two components in the total band. The contribution of the hard RXTE 
component to the ROSAT flux, however, seems to be negligible.

The detection of independent variability in the optical and the ROSAT regime 
favours multi-component models in order to explain the emission 
properties from the BL~Lac 0716+714. Single component models (e.\ g.\ simple 
SSC) are not capable of explaining the observed variability behaviour. The 
probable division of the RXTE band into two independently varying bands supports
this interpretation. From the data presented it is not clear how many components
contribute to the overall source flux; similar variability trends (Figure 
\ref{fg:lcrv_split}) in the optical and in the harder X-ray may help to limit
the number required.

  \section{Summary}
The preliminary analysis of the optical, ROSAT, and RXTE data obtained within a 
multifrequency campaign in March/April 1997 suggest complicated variability 
properties of the BL~Lac object 0716+714. The optical and the soft X-ray bands 
(ROSAT, soft RXTE) vary independently. In addition the hard RXTE band seems to 
vary independently from the soft X-ray band. Moreover, the hard RXTE band shows a flatter spectrum than the total RXTE band and reveals variability trends 
present in the optical data. These results provide evidence for a 
multi-component model of the source.  \\

{\small
{\noindent \bf Acknowledgment:} The work of K.\ O.\ and S.\ J.\ W.\ is
    supported by grants from DARA/DLR and DFG (SFB 328).
    The RXTE analysis was supported by NASA grant NAG5-3355, and
    M.\ J.\ H.\ acknowledges support from the PPARC.
    We thank the RXTE GOF members and Keith Jahoda for
    helpful advice. We also like to thank Holger Bock for kindly providing
    the optical data and Arno Witzel for useful discussions.
    }
  %
  
  %
  %
  %

\begin{thebibliography}{9}
    \bibitem{bie81} Biermann P., Duerbeck H., Eckart A.\ et al., 1981, ApJL, 
		    247, L53.
    \bibitem{bla78} Blandford R.\ D., Rees M.\ J. in: {\it Pittsburgh 
		      conference on BL~Lac Objects}, ed.\ A. N. Wolfe,
		      Pittsburgh, University of Pittsburgh Press, p.\ 328, 1978.
    \bibitem{chi97} Chiappetti L.,8th National Cosmic Physics (GIFCO) 
		    conference, 8 Apr 1997.\\
		    http://www.ifctr.mi.cnr.it/~lucio/WWW/Per\-sonal/sax.html
    \bibitem{den65} Dent W. A., 1965, Science, 148, 1458.
    \bibitem{eck86} Eckart A., Witzel A., Biermann P. et al., 1986, A\&A 168,
		    17.
    \bibitem{eck87} Eckart A., Witzel A., Biermann P. et al., 1987, A\&AS, 
		    67, 121.
    \bibitem{ede95} Edelson R., Krolik J., Madejski G. et al., 1995, ApJ, 438, 
		    120.
    \bibitem{kue81} K\"uhr H., Pauliny-Toth I. I. K., Witzel 
	     A., Schmidt J., 1981, AJ 86, 854.
    \bibitem{ree66} Rees M., 1966, MNRAS, 135, 345.
    \bibitem{sho65} Sholomitskii G. B., 1965, Sov.\ Astron., 9, 516.
    \bibitem{wag95} Wagner S.\ J., Witzel A., 1995, ARA\&A, 33, 163.
    \bibitem{wag96} Wagner S.\ J., Witzel A., Heidt J.\ et al., 1996, AJ, 111, 
		    2187.
    \bibitem{wol66} Woltjer L., 1966, ApJ, 146, 597.
  \end{thebibliography}
\end{document}